# Inverse operator representations of quantum phase

G.M.Saxena
Time and Frequency Section
National Physical Laboratory
Dr K.S.Krishnan Road New Delhi-110012 INDIA

E-mail: gmsaxena@mail.nplindia.ernet.in

**ABSTRACT**

We define quantum phase in terms of inverses of annihilation and creation operators. We show that like Susskind - Glogower phase operators, the measured phase operators and the unitary phase operators can be defined in terms of the inverse operators. However, for the unitary phase operator the Hilbert space includes the negative energy states. The quantum phase in inverse operator representation may find the applications in the field of quantum optics particularly in the squeezed states.

## 1. INTRODUCTION

The inverses of annihilation and creation operators were introduced [1] for obtaining the eigenvalue equation of the squeezed vacuum. These inverse operators were used in defining Susskind-Glogower phase operators [2,3] and eigen value equation of the squeezed coherent states [3]. In this paper we define the unitary and the measured phase operators in terms of these inverses of annihilation and creation operators. We briefly mention the properties of the annihilation and creation operators and their inverses before using them for defining the quantum phase operators.

The annihilation $a$ and creation $a^†$ operators for the bosons are defined by their actions on the number state $|n\rangle$ as follows

$$a|n\rangle = \sqrt{n}\,|n-1\rangle, \tag{1.1a}$$

$$a^†|n\rangle = \sqrt{(n+1)}\,|n+1\rangle \tag{1.1b}$$

and

$$N|n\rangle = a^† a\,|n\rangle = n|n\rangle. \tag{1.1c}$$

Similarly, the inverse annihilation and creation operators [1] are defined as

$$a^{-1}|n\rangle = |n+1\rangle/\sqrt{(n+1)} \tag{1.2a}$$

$$a^{†-1}|n\rangle = |n-1\rangle/\sqrt{n} \quad \text{for } n=1, 2 \ldots \tag{1.2b}$$

$$= 0 \quad \text{for } n=0. \tag{1.2c}$$

Here $a^{-1}$ and $a^{†-1}$ are the right and left inverses of $a$ and $a^†$ respectively i.e.,

$$a\,a^{-1} = a^{†-1} a^† = I \tag{1.3a}$$

$$a^{-1} a = a^† a^{†-1} = I - |0\rangle\langle 0|, \tag{1.3b}$$

here I is the identity operator and $|0\rangle\langle 0|$ is the projection operator on the vacuum. From eqs. (1.3) it follows that $a^{-1}$ and $a^{\dagger -1}$ are one-sided inverse operators. This property is used in defining Susskind-Glogower phase operators which are one-sided unitary.

We also introduce the following entities for positive number states

$$N^{1/2} |n\rangle = \sqrt{(n)} |n\rangle \qquad (1.4a)$$

and

$$N^{-1/2} |n\rangle = 1/\sqrt{(n)} |n\rangle \qquad (1.4b)$$
$$= 0 \text{ for } n=0$$

as these are useful in defining the inverse operator representations of quantum phase.

## 2. PHASE OPERATORS:

In this section we define the quantum phase operators, for the quantized electromagnetic field, in terms of the inverses of annihilation and creation operators. We have particularly chosen Susskind-Glogower, unitary and measured phase operators for obtaining their inverse operator representations. These phase operators are of interest due to the fact that they have non-classical nature and in the limiting case exhibit classical behaviour as well.

### 2.1 Susskind-Glogower Phase Operators

The properties of the inverses of annihilation and creation operators make them a natural choice for defining the Susskind-Glogower phase operators. We first consider the exponential Susskind-Glogower phase operators [4,5] which are defined as

$$e_s^{i\varphi} = \sum_{n=0}^{\infty} |n\rangle\langle n+1| \qquad (2.1)$$

$$e_s^{-i\varphi} = \sum_{n=0}^{\infty} |n+1\rangle\langle n| . \tag{2.2}$$

These operators satisfy the following relations:

$$e_s^{i\varphi} |n\rangle = |n-1\rangle, \tag{2.3a}$$

$$e_s^{-i\varphi} |n\rangle = |n+1\rangle, \tag{2.3b}$$

$$e_s^{i\varphi} |0\rangle = 0, \tag{2.4}$$

$$e_s^{i\varphi} e_s^{-i\varphi} = I, \tag{2.5}$$

and

$$e_s^{-i\varphi} e_s^{i\varphi} = I - |0\rangle\langle 0|. \tag{2.6}$$

We now define the exponential Susskind-Glogower phase operators in terms of the annihilation operator and its inverse i.e.,

$$e_s^{i\varphi} = aN^{-1/2} \tag{2.7}$$

and

$$e_s^{-i\varphi} = N^{1/2} a^{-1}, \tag{2.8}$$

here

$$N^{1/2} = \sum_{n=0}^{\infty} n^{1/2} |n\rangle\langle n| \tag{2.9}$$

and

$$N^{-1/2} = \sum_{n=0}^{\infty} n^{-1/2} |n\rangle\langle n|, \tag{2.10}$$

$$= 0 \quad \text{for } n=0$$

We observe that Susskind-Glogower phase operators defined in terms of the inverse operators satisfy the properties

$$e_s^{i\varphi} e_s^{-i\varphi} = aN^{-1/2} N^{1/2} a^{-1} = I \qquad (2.11)$$

and

$$e_s^{-i\varphi} e_s^{i\varphi} = N^{1/2} a^{-1} aN^{-1/2} = I - |0\rangle\langle 0| \qquad (2.12)$$

as (cf. eqn 1.3b)

$$a^{-1}a = I - |0\rangle\langle 0|.$$

Susskind - Glogower phase operators may also be defined in terms of the creation operator and its inverse i.e.,

$$e_s^{i\varphi} = a^{\dagger-1} N^{1/2} \qquad (2.13a)$$

and

$$e_s^{-i\varphi} = N^{1/2} a^{\dagger}. \qquad (2.13b)$$

It may be readily verified that

$$e_s^{i\varphi} e_s^{-i\varphi} = a^{\dagger-1} N^{1/2} N^{-1/2} a^{\dagger} = I \qquad (2.14a)$$

and

$$e_s^{-i\varphi} e_s^{i\varphi} = N^{-1/2} a^{\dagger} a^{\dagger-1} N^{1/2} = I - |0\rangle\langle 0|. \qquad (2.14b)$$

We observe that the inverses of annihilation and creation operators define the Susskind-Glogower phase operators in a simplified and straight forward manner. As the eigen operators of the squeezed vacuum and the squeezed coherent states involve the inverse of the creation and annihilation operators [1,3], the inverse operator representation of quantum phase may be useful in discussions of the phase of the squeezed states. These quantum phase operators, in the context of the squeezed states, may play significant role as the squeezed states have phase dependent noise fluctuations.

The cosine and sine of the phase play an important role in phase operator formalism and phase measurement. It may be readily verified that in the inverse operator representation of Susskind-Glogower phase operators, the number state expectation values of cosine and sine phase operators satisfy the relation

$$\langle n|(\cos_S\varphi)^2|n\rangle + \langle n|(\sin_S\varphi)^2|n\rangle = 1 \qquad (2.14c)$$

for $n \neq 0$. However, the vacuum expectation values of $\langle(\cos_S\varphi)^2\rangle$ and $\langle(\sin_S\varphi)^2\rangle$ are ¼ each instead of ½.

## 2.2 Unitary Phase Operators

The Susskind - Glogower and the unitary phase operators satisfy Lerner criterion [6]. This implies that in the positive energy domain, the unitary phase operator is well behaved like Susskind-Glogower phase operator. The similarity of both these operators in Hilbert space of positive number states ensures that even the unitary phase operators may be defined in terms of the inverse annihilation operators. However, for defining the unitary phase operators, the negative energy states are also included. Theoretically, the negative energy states are not forbidden but these states are decoupled from the positive energy states. We shall first briefly describe the existing unitary phase operators and then define them in terms of inverses of annihilation and creation operators. In these descriptions of unitary phase operators the negative energy states are included. The exponential unitary phase operators are defined [6] in terms of number states as

$$e_U^{i\varphi} = \sum_{n=-\infty}^{\infty} |n\rangle\langle n+1| \qquad (2.15a)$$

and

$$e_U^{-i\varphi} = \sum_{n=-\infty}^{\infty} |n+1\rangle\langle n|. \qquad (2.15b)$$

We now define these unitary phase operators in terms of the inverse annihilation and creation operators. The unitary exponential phase operators may be expressed in terms of the inverse of creation and annihilation operators, i.e.,

$$e_U^{i\varphi} = a^{\dagger-1} |N^{1/2}| \qquad (2.16a)$$

$$e_U^{-i\varphi} = |N^{1/2}| a^{-1}. \qquad (2.16b)$$

$|N^{1/2}|$ is the Hermitian amplitude operator defined as

$$|N^{1/2}| = \sum_{n=-\infty}^{\infty} |n^{1/2}| \, |n\rangle\langle n|. \qquad (2.17)$$

Here the number operator **N** and the identity operator I operate in the extended basis ($-\infty$ to $\infty$). They retain their properties valid in positive energy states in the extended basis ($-\infty$ to $\infty$) as well i.e.,

$$\mathbf{N}|n\rangle = \mathbf{n}|n\rangle \qquad (2.18a)$$

$$\sum_{n=-\infty}^{\infty} |n\rangle\langle n| = I. \qquad (2.18b)$$

It may be readily verified that in the inverse operator representation, the unitary phase operators satisfy the following relations

$$e_U^{i\varphi} e_U^{-i\varphi} = a^{\dagger-1} |N^{1/2}| |N^{1/2}| a^{-1} = I \qquad (2.19a)$$

and

$$e_U{}^{-i\varphi}\, e_U{}^{i\varphi} = |N^{1/2}|\, a^{-1}\, a^{\dagger-1}\, |N^{1/2}| = I. \qquad (2.19b)$$

From the above inverse operator representations of the unitary phase operator, it may be readily shown that the number state expectation value of cosine and sine phase operators fulfill the following relation

$$\langle n|(\cos_U\varphi)^2|n\rangle + \langle n|(\sin_U\varphi)^2|n\rangle = 1 \qquad (2.20)$$

for all real values of n. Here the Hilbert space includes both positive and negative number states and the number operator N operates on both positive and negative energy states.

## 2.3 Measured Phase Operator

The measured quantum phase operators which appear in homodyne techniques of measurement are supposed to reproduce the classical results as well. We now define the inverse operator representation of the exponential measured phase operators as

$$e_M{}^{i\varphi} = 2k a^{\dagger-1} \qquad (2.21a)$$

and

$$e_M{}^{-i\varphi} = 2k a^{-1}. \qquad (2.21b)$$

for the optical measurements. The measured phase operators are the normalized quadrature phase operators. We may define, using eqs.(2.21), the inverse operator representation of the quadrature phase operators as

$$\cos_M\varphi = k(a^{\dagger-1} + a^{-1}) \qquad (2.22a)$$

$$\sin_M\varphi = -ik(a^{\dagger-1} - a^{-1}) \qquad (2.22b)$$

here k is a state dependent c number such that $k = 1/2\,[\sqrt{n(n+1)}/(n+1/2)]$ with n being the number of photons in the field. It may be shown that in the inverse operator representation of the measured phase, the number state expectation values of cosine and sine phase operators or of the quadrature phase operators satisfy the relation

$\langle n|(\cos_M \varphi)^2|n\rangle + \langle n|(\sin_M \varphi)^2|n\rangle = 1$,  (2.23)

for positive values of n including vacuum state. In the case of the measured phase operator, the Hilbert space includes only positive energy states and the number operator N spans positive number states only unlike the unitary phase operators in which Hilbert space includes negative energy or number states as well.

**CONCLUSION**

In this paper we have introduced inverse creation and annihilation operators representations of the Susskind-Glogower, unitary and measured phase operators. We find that inverses of annihilation and creation operators are the natural choice for defining the Susskind-Glogower and the unitary phase operators. The number state expectation values of cosine and sine phase operators in inverse operator representation of unitary and measured phase operators, defined in the paper, readily satisfy the trigonometric identity. The inverse annihilation operators were introduced by us for defining the eigenvalue equations of the squeezed vacuum and the squeezed coherent states. It is significant to mention that the eigenvalue equations of the squeezed coherent states [3], defined using inverse of annihilation and creation operators, are consistent unlike the eigen value equation of the squeezed coherent state defined by Yuen [7] which can not be reduced to the squeezed vacuum on putting the coherent amplitude equal to zero. This fact has highlighted the importance of the inverse of annihilation and creation operators in the field of quantum optics. The application of the inverses of annihilation and creation operators to defining the quantum phase may be useful in the discussions of the quantum phase in quantum optics and particularly in the squeezed states as these states have phase dependent quantum noise and are the eigenstates of the operator involving inverse annihilation operators.